\documentclass[journal]{IEEEtran}
\ifCLASSINFOpdf
   \usepackage[pdftex]{graphicx}
\else
\fi
\usepackage{mathtools}
\usepackage{amsmath,color,amsthm}
\usepackage{breqn}
\usepackage{bm}
\usepackage{cite}
\usepackage{algorithm,algpseudocode}
\usepackage{caption}
\usepackage{lipsum}
\usepackage{cuted}
\usepackage{enumerate}
\usepackage{booktabs} 
\usepackage{graphicx}
\usepackage{float}
\usepackage{mathrsfs}

\usepackage{amsmath,amsthm,amssymb}

\makeatletter
\newcommand{\mathleft}{\@fleqntrue\@mathmargin0pt}
\newcommand{\mathcenter}{\@fleqnfalse}
\makeatother

\DeclarePairedDelimiterX\Basics[1](){ #1}

\usepackage{array}

\usepackage{stfloats}
\usepackage{url}

\begin{document}
%
\title{Underwater Optical Communication System Relayed by $\alpha-\mu$ Fading Channel: Outage, Capacity and Asymptotic Analysis }
%
%
%

\author{Mohammed Amer and Yasser Al-Eryani\thanks{M. Amer is with the Department of Electrical Engineering at King Fahd University of Petroleum and Minerals (KFUPM), Dhahran, KSA (Emails: M\_Amer@mail.com). Yasser Al-Eryani is with the Department of Electrical and Computer Engineering at the University of Manitoba, Canada (Emails: Yasser.Aleryani@umanitoba.ca).}}
\maketitle

\begin{abstract}
We investigate underwater optical communication system that is relayed by a single decode-and-forward (DF) relay through an exponential-generalized Gamma distribution (EGG) into a final destination.
Specifically, a certain terminal device sends data through underwater wireless optical link (UWO) that utilizes the so-called blue laser technology into a nearby relay that in term sends a decoded (and modulated) version of the received signal into a remote destination. 
The RF link is assumed to follow the generalized $\alpha-\mu$ distribution; which include many distributions as a special cases, e.g., Rayleigh. In the other hand, the UWO link is presumed to follow the state-of-art Exponential-Generalized Gamma distribution (EGG) which was recently proposed to model the underwater optical turbulence. Closed-form expressions of outage probability, average error rate and ergodic capacity are derived assuming heterodyne detection technique (HD). Also, asymptotic outage expression is obtained for more performance insights. 
Results show that high achievable rate is obtained for high-speed underwater communication systems when turbulence conditions underwater are relatively weak. In addition, the RF link is dominating the outage performance in weak optical turbulence while UWO link is dominating the outage performance in severe optical turbulence.
\end{abstract}

\begin{IEEEkeywords}
underwater communication; DF relaying; unified EGG; $\alpha-\mu$ fading, performance analysis.
\end{IEEEkeywords}

%
\IEEEpeerreviewmaketitle

\section{Introduction}

\IEEEPARstart{R}{ecently}, underwater wireless communication (UWC) has attracted lot of research attention for the wide range of underwater applications such as offshore seismic surveys, seafloor monitoring, submarine navigation, and military defense activities. 
In general, UWC suffers from many obstacles that effect the underwater signal propagation for long distances such as scattering (due to large water particles compared to free space), turbulence, and absorption phenomena. 
Such effects are caused by the transmission of the signal through an unguided variant water environments~\cite{1}. 
In its current status, most of UWC systems are implemented using both RF and acoustic carriers where they are suffering from the high latency, low data rates, and band limitation.
Such low latency and unsatisfactory data rates severely contradicts with future 5G and beyond 5G (B5G) applications such as underwater traffic between coastal cities.
Accordingly, underwater wireless optical communication (UWOC) is proposed as a promising technology for the large data-rates (Gbps levels), high security and bandwidth~\cite{2}. \par
In literature, different studies of the transmission of optical information-bearing signals throughout water (salty and fresh) have been conducted theoretically and experimentally. In~\cite{3}-\cite{4}, the authors characterized the UWOC mathematical channel model using radiative transfer and back-reflection theories and then, investigated the performance analysis based on the estimated channel effects. Furthermore, the performance analysis of hybrid optical/acoustic communication system is proposed and studied in~\cite{5} while multi-hop Decode-and-Forward (DF) is investigated in~\cite{6}.\par
Nevertheless, all previous works within the literature has assumed the UWC channel fading effect to follow log-normal distribution which does not include the underwater turbulence and only approximately estimate the scattering effects of salty waters on propagating optical waves~\cite{7}\footnote{Underwater turbulence is mainly related to the temperature fluctuations, salinity variations, and the existence of air bubbles in seawater caused by quick transition of the water refractive index that influence the optical signals~\cite{8}.}.\par
Another thing to concern about when studying the practicality of UWOC is that they only support small distances due to the exponential degradation of the signal strength versus physical underwater distance.
Accordingly, the existence of some relaying mechanism that first receive the underwater optical signal from the closes free space point and then relay it to its final destination.
In this work, we investigate the performance analysis of UWOC link that is relayed by an $\alpha-\mu$ RF channel into a final destination. 
To the best of author's knowledge, the performance analysis of one-way mixed underwater optical communication (UOC)/RF relaying has not been investigated or analyzed yet. 
The major contributions of this article can be summarized as follows:
\begin{itemize}
\item We propose and evaluating the performance analysis of OW mixed UOC/RF relaying using unified Exponential-Generalized Gamma (EGG) statistical channel model with underwater optical turbulence impairments and generalized $\alpha-\mu$ channel.
\item We derive a closed-form expressions for the probability of outage, average symbol error rate and ergodic channel capacity of the proposed system model.
\item Obtain the asymptotic outage probability (At high SNR) in order to have deep insights about the impact of UOT on the overall outage performance.
\item Analyze the influence of air bubbles, under-water optical turbulence, water type, water temperature, and $\alpha-\mu$ parameters on the overall performance of the system.
\end{itemize}\par

The rest of this paper is organized as follows. In Section II proposes the system model and its associated channel models as well as the CDF of each link. The exact analysis of outage probability of the system is derived in Section III. In Section IV, we obtain the asymptotic expression of the outage probability formula. Section V obtains the ASEP in closed-form formula. In Section VI, the ergodic capacity is derived in closed-form expression. the discussions of various numerical and simulation results is in Section VII. Finally, concluding the work is given in Section VIII.

\section{System and Channel Model}

\begin{figure}[!b]
	\centering
	\includegraphics[height=7cm, width=8.7cm]{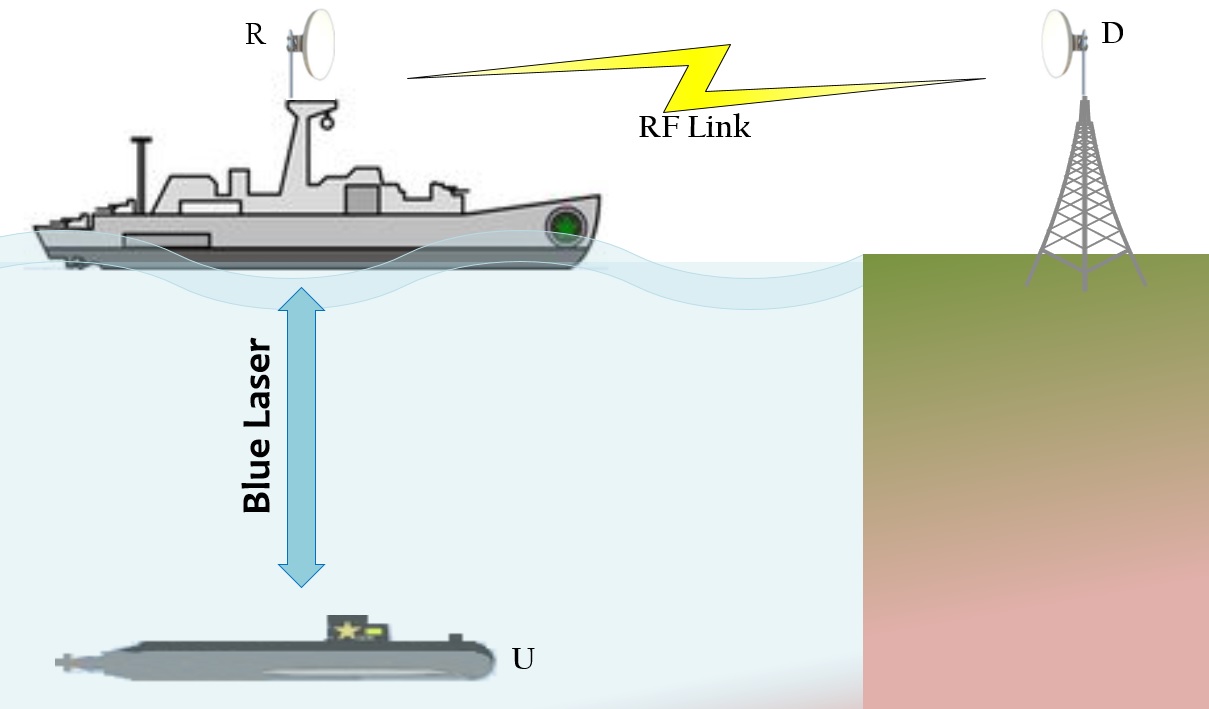}
	\caption{Example scenario for underwater optical communication network. }\label{Fig.1}
\end{figure} 
Consider a dual-hop mixed UOC/RF relay network composed of a source node (U) on the first hop, one un-coded DF relay (R), and destination node on the second hop (D) as shown in Fig. 1. The source is presumed to communicate with relay node using UWO link; this relay forward the data to the destination node through RF link. the user node is assumed to be equipped with a single photo-aperture transmitter while the relay is equipped with a single photo-aperture detector and a single transmitter antenna, and the destination node is equipped with a single receiver antenna. Moreover, the direct link between the source and destination nodes is presumed to be in deep fade thus, it is not carried in the analysis of this paper. The communication type between the $U\to R$ and $R\to D$ links is operated in half-duplex mode and performed in two phases: $U\to R$ and $R\to D$. The received optical signal at the input of $R$ from the $U$ is expressed by:
\begin{equation}
\begin{aligned}
r_{R}^{Opt}=g_{R,D}\bigg\{\sqrt{P_{U}^{Opt}}(1+M x_{U})\bigg\}+n_{R}^{Opt},
\end{aligned}
\end{equation}
where $P_{U}^{Opt}$ is the average transmitted optical power; which relate to the electrical power $(P)$ by the electrical-to-optical conversion ratio $\zeta$ as $P=\zeta\times P^{Opt}$. $M$ is the modulation index, and $x_{U}$ is the transmitted symbols of $U$ with $\mathbb{E}\big[|x_{U}|^{2}\big]=1$, where $\mathbb{E}[\cdot]$ is the expectation notation. $g_{R,D}$ is the small-scale channel coefficients of the $U\to R$ link, and $n_{R}^{Opt}$ is the zero-mean additive white Gaussian noise (AWGN) with power spectral density (PSD) of $N_{0,R}^{Opt}$. The instant. SNR at the input of R is given by
\begin{equation}
\begin{aligned}
\gamma_{R}=\frac{P_{U}\zeta}{N_{0,R}^{Opt}} |g_{R,D}|^2.
\end{aligned}
\end{equation}
The received RF signal at the input of $D$ from $R$ in the second phase is given as
\begin{equation}
\begin{aligned}
r_{D}^{RF}=\sqrt{P_{R}^{RF}}h_{R,D}x_{R}+n_{D}^{RF},
\end{aligned}
\end{equation}
where $P_{R}^{RF}$ and $x_{R}$ are the transmitted electrical power and symbol of $R$, respectively. $h_{R,D}$ is the small-scale channel coefficient of the $R\to D$ link. $n_{D}^{RF}$ is the zero-mean AWGN with PSD of $N_{0,D}^{RF}$. The instant. SNR at the input of $D$ is
\begin{equation}
\begin{aligned}
\gamma_{D}=\frac{P_{R}}{N_{0,D}^{RF}} |h_{R,D}|^2.
\end{aligned}
\end{equation}

\subsection{RF Channel Model}\par
In the $U\to R$ link, the channel coefficient $|h_{R,D}|$ is following the generic $\alpha-\mu$ fading model. Therefore, the channel gains $|g_{R,D}|^{2}$ probability density function (PDF) is given by~\cite{9}
\begin{equation}
\begin{aligned}
f_{\gamma_{R,D}}(\gamma_{R,D})=\frac{\alpha}{2\Gamma{[\mu]}}{\frac{\mu^{\mu}}{(\bar{\gamma}_{R,D})^{\frac{\alpha\mu}{2}}}}\bigg({\gamma_{R,D}}\bigg)^{\frac{\alpha\mu}{2}-1}e^{-\mu\big({\frac{\gamma_{R,D}}{\bar{\gamma}_{R,D}}}\big)^{\frac{\alpha}{2}}},
\end{aligned}
\end{equation}
where ${\mu\geq0}, {\alpha\geq0},  {\gamma_{R,D}\geq0}$, $\Gamma[\cdot]$ is the generalized gamma function defined in~\cite{10}. The parameters $\alpha$ and $\mu$ are used to model non-linearity and multi-path propagations through random medium and $\bar{\gamma}_{R,D}=\mathbb{E}\{\gamma_{R,D}\}=\frac{P_{R}}{\mathbb{N}_{0,D}^{RF}}\mathbb{E}\{|h_{R,D}|^2\}$ is the average received SNR. The model is generalized to other fading distributions such as Rayleigh, Exponential, Weibull, Nakagami-m, and one-sided Gaussian by changing the parameters value of $\alpha$ and $\mu$. The cumulative distribution function (CDF) of the $\alpha-\mu$ is obtained by: $\int_{0}^{\gamma_{R,D}}f_{\gamma_{\varsigma}}(\varsigma)d\varsigma$, given in terms of Meijer's G-Function as~\cite{9}
\begin{equation}
\begin{aligned}
F_{\gamma_{R,D}}{(\gamma_{R,D})}=\frac{1}{\Gamma{[\mu]}}G_{1,2}^{1,1}\left[\mu({\frac{\gamma_{R,D}}{\bar{\gamma}_{R,D}}})^{\frac{\alpha}{2}}\left|\begin{matrix}1\\
\mu,0\end{matrix}\right.\right],
\end{aligned}
\end{equation}
where $\large{G}_{c,d}^{a,b} \left[ \cdot  \left|
\begin{smallmatrix} .,. \\ .,. \end{smallmatrix} \right. \right]$ is the Meijer's G-function defined in~\cite{10}.

\begin{table}[!t]
\small\addtolength{\tabcolsep}{-4.0pt}
\caption{System parameters of a single UWO link.}
\label{Table.1}
\centering
\begin{tabular}{ l c c c c c c c} 
\hline
W. Type & Turb. & BL & a & b & c & $\lambda$ & w\\ [0.5ex]
\hline
Salty & Weak & 2.4 & 0.7736 & 1.1372 & 49.1773 & 0.4687 & 0.1770\\
Salty & Moderate & 4.7 & 0.5307 & 1.2154 & 35.7368 & 0.3953 & 0.2064 \\
Salty & Severe & 16.5 & 0.0161 & 3.2033 & 82.1030 & 0.1368 & 0.4951\\
Fresh & Weak  & 2.4 & 3.7291 & 1.0721 & 30.3214 & 0.5273 & 0.1953\\
Fresh & Moderate & 4.7 & 1.2526 & 1.1501 & 41.3258 & 0.4603 & 0.2109\\
Fresh & Severe & 16.5 & 0.0075 & 2.9963 & 216.8356 & 0.1602 & 0.5117\\
\hline
\end{tabular}
\end{table}


\subsection{UWO Channel Model}\par
The underwater optical channel of the $U\to R$ link is assumed to experience the unified Exponential-Generalized Gamma (EGG) model with underwater optical turbulence impairments. Under Heterodyne Detection ($r=1$), the PDF of $\gamma_{U,R}$ is written as~\cite{11}
\begin{equation}
\begin{aligned}
f_{\gamma_{U,R}}(\gamma_{U,R})&=\frac{w}{\lambda\gamma_{U,R}}e^{-\frac{\gamma_{U,R}}{\lambda\bar{\gamma}_{U,R}}}\\
&+\frac{c(1-w)}{\Gamma{[a]}\gamma_{U,R}}G_{0,1}^{1,0}\left[\bigg({\frac{\gamma_{U,R}}{b\bar{\gamma}_{U,R}}}\bigg)^{c}\left|\begin{matrix}-\\
a\end{matrix}\right.\right],
\end{aligned}
\end{equation}
where $\bar{\gamma}_{U,R}=\mathbb{E}\{\gamma_{U,R}\}=\frac{P_{U}\zeta}{\mathbb{N}_{0,R}^{Opt}}\mathbb{E}\{|g_{U,R}|^2\}$; $a$, $b$, and $c$ are the fading parameters related to the Generalized-Gamma distribution which characterize the water Salinity and air Bubble levels (BL) impairments. $\lambda$ is the Exponential distribution parameter and $0< w<1$. Table. I illustrates the different UWO numerical values of each parameter and the corresponding UOT scenarios used in this work.\par
The CDF $F_{\gamma_{U,R}}(\gamma_{U,R})$ of a single UWO link is obtained by integrating the PDF in Eq. (7) with respect to $\gamma_{U,R}$, and it is given by 
\begin{dmath}
\begin{aligned}
F_{\gamma_{U,R}}(\gamma_{U,R})&=wG_{1,2}^{1,1}\left[\frac{1}{\lambda}\bigg({\frac{\gamma_{U,R}}{\bar{\gamma}_{U,R}}}\bigg)\left|\begin{matrix}1\\
1,0\end{matrix}\right.\right]+\frac{(1-w)}{\Gamma{[a]}}\\
&\times G_{1,2}^{1,1}\left[\frac{1}{b^{c}}\bigg({\frac{\gamma_{U,R}}{\bar{\gamma}_{U,R}}}\bigg)^{c}\left|\begin{matrix}1\\
a,0\end{matrix}\right.\right].
\end{aligned}
\end{dmath}

\section{Exact Outage Probability Analysis}

The outage performance is a critical metric in wireless systems which define as the probability that the instant. $\gamma_{X,Y}$ falls below a predetermined threshold value $\gamma_{out}$, mathematically seen as $P_{out}=Pr{(\gamma\leq\gamma_{out})}$; where $P[\cdot]$ is the probability notation. The end-to-end outage probability, assuming independent and identical distribution (i.i.d.), is given by
\begin{equation}
\begin{aligned}
P_{out}= F_{\gamma_{U,R}}^{Opt}{(\gamma_{out})}+F_{\gamma_{R,D}}^{RF}{(\gamma_{out})}- F_{\gamma_{U,R}}^{RF}{(\gamma_{out})}F_{\gamma_{R,D}}^{Opt}{(\gamma_{out})},
\end{aligned}
\end{equation}
where $F_{\gamma_{U,R}}^{Opt}{(\gamma_{out})}$ and $F_{\gamma_{R,D}}^{RF}{(\gamma_{out})}$ are the CDFs of the first and second hops, respectively. By substitution Eq. (6) and Eq. (8) into Eq. (9) with a straightforward manipulation and simplification, the probability of outage of the proposed system is then given by Eq. \ref{Outage} at the top of this page, where $G_{.,.:.,.:.,.}^{.,.:.,.:.,.}\left[\begin{smallmatrix}.\cr .\end{smallmatrix}\Bigg\vert\begin{smallmatrix}\ .,.\cr .,.\end{smallmatrix}\Bigg\vert\begin{smallmatrix}\ .,.\cr .,.\end{smallmatrix}\Bigg\vert \psi,\chi \right]$ is the Extended Generalized Bivariate Meijer's G-Function (EGBMGF)~\cite{12}.
\begin{figure*}[!t]
	\normalsize
	\begin{dmath}\begin{aligned}\label{Outage}
P_{out}&= \frac{1}{\Gamma{[\mu]}}G_{1,2}^{1,1}\left[\mu\bigg({\frac{\gamma_{out}}{\bar{\gamma}_{R,D}}}\bigg)^{\frac{\alpha}{2}}\left|\begin{matrix}1\\
	\mu,0\end{matrix}\right.\right]+wG_{1,2}^{1,1}\left[\frac{1}{\lambda}\bigg({\frac{\gamma_{out}}{\bar{\gamma}_{U,R}}}\bigg)\left|\begin{matrix}1\\
	1,0\end{matrix}\right.\right]+\frac{(1-w)}{\Gamma{[a]}}\\
&\times G_{1,2}^{1,1}\left[\frac{1}{b^{c}}\bigg({\frac{\gamma_{out}}{\bar{\gamma}_{U,R}}}\bigg)^{c}\left|\begin{matrix}1\\
	a,0\end{matrix}\right.\right]-\frac{w}{\Gamma{[\mu]}} G_{0,0:1,2:1,2}^{0,0:1,1:1,1}\left[\begin{matrix}-\cr -\end{matrix}\Bigg\vert\begin{matrix}1\cr 1,0\end{matrix}\Bigg\vert\begin{matrix}1\cr \mu,0\end{matrix}\Bigg\vert {\frac{1}{\lambda}\bigg({\frac{\gamma_{out}}{\bar{\gamma}_{U,R}}}\bigg)},{\mu\bigg({\frac{\gamma_{out}}{\bar{\gamma}_{R,D}}}\bigg)^{\frac{\alpha}{2}}}\right]\\
&-\frac{(1-w)}{\Gamma{[a]}\Gamma{[\mu]}}G_{0,0:1,2:1,2}^{0,0:1,1:1,1}\left[\begin{matrix}-\cr -\end{matrix}\Bigg\vert\begin{matrix}1\cr a,0\end{matrix}\Bigg\vert\begin{matrix}1\cr \mu,0\end{matrix}\Bigg\vert {\frac{1}{b^{c}}\bigg({\frac{\gamma_{out}}{\bar{\gamma}_{U,R}}}\bigg)^{c}},{\mu\bigg({\frac{\gamma_{out}}{\bar{\gamma}_{R,D}}}\bigg)^{\frac{\alpha}{2}}}\right],
	\end{aligned}\end{dmath}
	\hrulefill
	\vspace*{4pt}
\end{figure*}

\section{Asymptotic Analysis of Outage Probability}
Due to the complex expression of the end-to-end outage probability of the system model, the impact of each parameter is ambiguous. Hence, Asymptotic Analysis shows more insights of the impact of various system parameters on the overall outage performance. The end-to-end outage expression (at high SNR regime) can be expressed as $P_{out}^{\bar{\gamma}\to\infty}\simeq G_{c}({SNR})^{-G_{d}}$, where $G_{c}$ and $G_{d}$ are the coding and diversity gains, respectively~\cite{13}. By assuming i.i.d case, that is, $\bar{\gamma}_{U,R}=\bar{\gamma}_{R,D}=\bar{\gamma}$, we can write the end-to-end asymptotic outage expression as the sum of each asymptotically individual CDF because the multiplication of two or more CDFs is a very small value and thus, we may ignore it. Finally, the asymptotic end-to-end outage can be shown as:
\begin{equation}
\begin{aligned}
P_{out}^{\bar{\gamma}\to\infty}\simeq F_{\gamma_{U,R}}^{Opt\to\infty}{(\gamma_{out})}+F_{\gamma_{R,D}}^{RF\to\infty}{(\gamma_{out})},
\end{aligned}
\end{equation}
where $F_{\gamma_{U,R}}^{Opt\to\infty}{(\gamma_{out})}$ and $F_{\gamma_{R,D}}^{RF\to\infty}{(\gamma_{out})}$ are the CDFs of each hop at high SNR. Starting by $F_{\gamma_{R,D}}^{RF\to\infty}{(\gamma_{out})}$, the asymptotic expression can be obtained by utilizing the generalized incomplete gamma function expansion series as
\begin{equation}
\begin{aligned}
F_{\gamma_{R,D}}^{RF\to\infty}{(\gamma_{out})}\simeq \frac{\Gamma{[\mu,0]}}{\Gamma{[\mu]}}+\frac{(\frac{\gamma_{out}}{\bar{\gamma}})^{\frac{\alpha\mu}{2}}}{\mu\Gamma{[\mu]}}-1,
\end{aligned}
\end{equation}
where $\Gamma{[\mu,0]}$ is the generalized incomplete gamma function~\cite{10}. \par
The asymptotic expression of $F_{\gamma_{U,R}}^{Opt\to\infty}{(\gamma_{out})}$ is given by 
\begin{equation}
\begin{aligned}
F_{\gamma_{U,R}}^{Opt\to\infty}{(\gamma_{out})}\simeq \frac{w\gamma_{out}}{\lambda\bar{\gamma}}+\frac{(1-w)}{\Gamma{[a+1]}}\bigg(\frac{\gamma_{out}}{b\bar{\gamma}}\bigg)^{ac}.
\end{aligned}
\end{equation}
Upon substituting Eq. (12) and Eq. (13) into Eq. (11), we derive the end-to-end asymptotic outage expression by
\begin{equation}
\begin{aligned}
P_{out}^{\bar{\gamma}\to\infty}&=\frac{w\gamma_{out}}{\lambda\bar{\gamma}}+\frac{(1-w)}{\Gamma{[a+1]}}\bigg(\frac{\gamma_{out}}{b\bar{\gamma}}\bigg)^{ac}+\frac{\Gamma{[\mu,0]}}{\Gamma{[\mu]}}\\
&+\frac{(\frac{\gamma_{out}}{\bar{\gamma}})^{\frac{\alpha\mu}{2}}}{\mu\Gamma{[\mu]}}-1.
\end{aligned}
\end{equation}
Rewriting Eq. (14) in the approximated-form while ignoring the small terms, we obtain
\begin{equation}
\begin{aligned}
P_{out}^{\bar{\gamma}\to\infty}=\bigg(\frac{\lambda\bar{\gamma}}{w\gamma_{out}}\bigg)^{-1}+\bigg(\Psi_{1}\frac{\bar{\gamma}}{\gamma_{out}}\bigg)^{-(ac)}+\bigg(\Psi_{2}{{\frac{\bar\gamma}{\gamma_{out}}\bigg)^{-(\frac{\alpha\mu}{2})}}},
\end{aligned}
\end{equation}
where $\Psi_{1}=\frac{b\Gamma{[a+1]}}{(1-w)}$ and $\Psi_{2}=\big({\mu\Gamma{[\mu]}\big)}^{-\frac{2}{\alpha\mu}}$. In Table II, the coding and diversity gains $(G_{c},G_{d})$ of the system is shown for different domination scenarios; whereas the notation $T$ represents the term number in Eq. (15).
\begin{table}[!b]
\caption{Coding gain and diversity order of the system model.}
\label{Table.3}
\centering
\small\addtolength{\tabcolsep}{0.7pt}
\begin{tabular}{ c|| c c } 
\hline
Domination link/s & $G_{d}$ & $G_{c}$ \\  [0.5ex]
\hline
$T_{1}$  & $1$ & $\frac{\lambda}{w\gamma_{out}}$\\ 

$T_{2}$  & $a\times c$ & $\frac{\Psi_{1}}{\gamma_{out}}$\\

$T_{3}$  & $\frac{\alpha\mu}{2}$ & $\frac{\Psi_{2}}{\gamma_{out}}$ \\

$T_{1}$ and $T_{3}$  & $\frac{\alpha\mu}{2}\simeq1$ & $\frac{\lambda}{w\gamma_{out}}+\frac{\Psi_{2}}{\gamma_{out}}$\\

$T_{1}$ and $T_{2}$  & $a\times c\simeq 1$ & $\frac{\lambda}{w\gamma_{out}}+\frac{\Psi_{1}}{\gamma_{out}}$\\

$T_{2}$ and $T_{3}$  & $\frac{\alpha\mu}{2}\simeq a\times c$ & $\frac{\Psi_{2}}{\gamma_{out}}+\frac{\Psi_{1}}{\gamma_{out}}$\\

$T_{1}$, $T_{2}$ and $T_{3}$  & $a\times c\simeq\frac{\alpha\mu}{2}\simeq1$ & $\frac{\Psi_{1}}{\gamma_{out}}+\frac{\Psi_{2}}{\gamma_{out}}+\frac{\lambda}{w\gamma_{out}}$\\
\hline
\end{tabular}
\end{table}

\section{Average Symbol Error Probability Analysis}
The ASEP is described as the average number of incorrectly received symbols as a result of bad channel quality. The end-to-end ASEP of the system model is expressed as
\begin{equation}
\begin{aligned}
ASEP_{e2e}=P_{L,1}^{Opt}+P_{L,2}^{RF}-2P_{L,1}^{Opt}P_{L,2}^{RF},
\end{aligned}
\end{equation}
where $P_{L,1}^{Opt}$ and $P_{L,2}^{RF}$ are the ASEP of each link, respectively. In general, the ASEP can be derived using the CDF-based approach by~\cite{14}
\begin{equation}
\begin{aligned}
P_{L,i}= \frac{\eta\sqrt{\beta}}{2\sqrt{\pi}}\int_{0}^{\infty}\frac{e^{-\beta\gamma}}{\sqrt{\gamma}}F_{\gamma}^{(i)}{(\gamma)}d\gamma\quad i=1,2,
\end{aligned}
\end{equation}
where $(\eta,\beta)>0$ are the modulation scheme parameters, e.g. BPSK $(\eta=\beta=1)$. Starting by the UWO link, the $P_{L,1}^{Opt}$ can be derived in closed-form expression by using the Fox's H-function and then utilizing Ref.~\cite{15} as
\begin{dmath}
P_{L,1}^{Opt}=\frac{\eta\sqrt{\beta}}{2\sqrt{\pi}}\bigg(\frac{w}{\sqrt{\beta}}H_{2,2}^{1,2}\left[{\frac{1}{\beta\lambda\bar{\gamma}}}\left|\begin{matrix}(\frac{1}{2},1)(1,1)\\
(1,1)(0,1)\end{matrix}\right.\right]+\frac{(1-w)}{\Gamma{[a]}\sqrt{\beta}}\\
\times H_{2,2}^{1,2}\left[\bigg({\frac{1}{\beta b\bar{\gamma}}}\bigg)^{c}\left|\begin{matrix}(\frac{1}{2},c)(1,1)\\
(a,1)(0,1)\end{matrix}\right.\right]\bigg),
\end{dmath}
where $\large{H}_{c,d}^{a,b} \left[ \cdot  \left|
\begin{smallmatrix} .,. \\ .,. \end{smallmatrix} \right. \right]$ is the Fox's H-function defined in~\cite{16}. The $P_{L,2}^{RF}$ can be derived in closed-form by using Ref.~\cite{15} as
\begin{dmath}
P_{L,2}^{RF}=\frac{\eta\sqrt{\beta}}{2\sqrt{\pi}}\bigg(\frac{1}{\sqrt{\beta}\Gamma{[\mu]}}H_{2,2}^{1,2}\left[\mu\bigg(\frac{1}{\beta\bar{\gamma}}\bigg)^{\frac{\alpha}{2}}\left|\begin{matrix}(\frac{1}{2},\frac{\alpha}{2})(1,1)\\
(\mu,1)(0,1)\end{matrix}\right.\right]\bigg).
\end{dmath}
Finally, by substituting Eqs. (18) and (19) into Eq. (16) we get the end-to-end ASEP is given by Eq. \ref{ASEP}.
\begin{figure*}[!t]
	\normalsize
	\begin{dmath}\begin{aligned}\label{ASEP}
ASEP_{e2e}&=\frac{\eta\sqrt{\beta}}{2\sqrt{\pi}}\Bigg\{\frac{w}{\sqrt{\beta}}H_{2,2}^{1,2}\left[{\frac{1}{\beta\lambda\bar{\gamma}}}\left|\begin{matrix}(\frac{1}{2},1)(1,1)\\
(1,1)(0,1)\end{matrix}\right.\right]+\frac{(1-w)}{\Gamma{[a]}\sqrt{\beta}}H_{2,2}^{1,2}\left[\bigg({\frac{1}{\beta b\bar{\gamma}}}\bigg)^{c}\left|\begin{matrix}(\frac{1}{2},c)(1,1)\\
(a,1)(0,1)\end{matrix}\right.\right]\\&
+\frac{1}{\sqrt{b}\Gamma{[\mu]}}H_{2,2}^{1,2}\left[\mu\bigg(\frac{1}{b\bar{\gamma}}\bigg)^{\frac{\alpha}{2}}\left|\begin{matrix}(\frac{1}{2},\frac{\alpha}{2})(1,1)\\
(\mu,1)(0,1)\end{matrix}\right.\right]-\frac{2}{\sqrt{\beta}\Gamma{[\mu]}}H_{2,2}^{1,2}\left[\mu\bigg(\frac{1}{\beta\bar{\gamma}}\bigg)^{\frac{\alpha}{2}}\left|\begin{matrix}(\frac{1}{2},\frac{\alpha}{2})(1,1)\\
(\mu,1)(0,1)\end{matrix}\right.\right]\\
&\times\Bigg(\frac{w}{\sqrt{\beta}}H_{2,2}^{1,2}\left[{\frac{1}{\beta\lambda\bar{\gamma}}}\left|\begin{matrix}(\frac{1}{2},1)(1,1)\\
(1,1)(0,1)\end{matrix}\right.\right]+\frac{(1-w)}{\Gamma{[a]}\sqrt{\beta}}H_{2,2}^{1,2}\left[\bigg({\frac{1}{\beta b\bar{\gamma}}}\bigg)^{c}\left|\begin{matrix}(\frac{1}{2},c)(1,1)\\
(a,1)(0,1)\end{matrix}\right.\right]\Bigg)\Bigg\}.
	\end{aligned}\end{dmath}
	\hrulefill
	\vspace*{4pt}
\end{figure*}

\section{Ergodic Capacity}
The ergodic capacity ($C$), that is also well-known as the achievable rate, is an important metric to quantify the max. transmission rate of under-water optical/RF communication system. Generally, the ergodic capacity can be derived using the following expression:
\begin{equation}
\begin{aligned}
C&=\int_{0}^{\infty}\log_{2}{(1+\gamma)}f_{\gamma}(\gamma)d\gamma,\\
&=\frac{1}{\ln{(2)}}\int_{0}^{\infty}\ln{(1+\gamma)}f_{\gamma}(\gamma)d\gamma~~~\text{bps/Hz}.
\end{aligned}
\end{equation}
To evaluate the ergodic capacity in Eq. (21) we need first to derive the end-to-end PDF by $f_{\gamma_{e2e}}(\gamma)=\frac{d}{d\gamma}P_{out}$. Upon using [Eq. (2.9.1)] as well [Eq. (2.2.1)] in Ref.~\cite{16}, then the pdf $f_{\gamma_{e2e}}\left(
\gamma\right)$ will be given by Eq. (\ref{fe2e}) at the top of next page.

\begin{figure*}[!t]
	\normalsize
	\begin{dmath}\label{fe2e}
f_{\gamma_{e2e}}(\gamma)=\frac{1}{\gamma\Gamma{[\mu]}}H_{2,3}^{1,2}\left[\bigg(\frac{\mu\gamma}{\bar{\gamma}_{R,D}}\bigg)\left|\begin{matrix}(0,1)(1,1)\\
	(\mu,1)(0,1)(1,1)\end{matrix}\right.\right]+\frac{w}{\gamma}H_{2,3}^{1,2}\left[\bigg(\frac{\gamma}{\lambda\bar{\gamma}_{U,R}}\bigg)\left|\begin{matrix}(0,1)(1,1)\\
	(1,1)(0,1)(1,1)\end{matrix}\right.\right]\\
+\frac{(1-w)}{\gamma\Gamma{[a]}}H_{2,3}^{1,2}\left[\bigg(\frac{\gamma}{b\bar{\gamma}_{U,R}}\bigg)^{c}\left|\begin{matrix}(0,1)(1,1)\\
	(a,1)(0,1)(1,c)\end{matrix}\right.\right]-\frac{w}{\Gamma{[\mu]}}\Bigg(H_{1,2}^{1,1}\left[\bigg({\frac{\gamma}{\lambda\bar{\gamma}_{U,R}}}\bigg)\left|\begin{matrix}(1,1)\\
	(1,1)(0,1)\end{matrix}\right.\right]\\
\times \frac{1}{\gamma}H_{2,3}^{1,2}\left[\bigg(\frac{\mu\gamma}{\bar{\gamma}_{R,D}}\bigg)\left|\begin{matrix}(0,1)(1,1)\\
	(\mu,1)(0,1)(1,1)\end{matrix}\right.\right]+H_{1,2}^{1,1}\left[\bigg(\frac{\mu\gamma}{\bar{\gamma}_{R,D}}\bigg)\left|\begin{matrix}(1,1)\\
	(\mu,1)(0,1)\end{matrix}\right.\right]\\
\times \frac{1}{\gamma}H_{2,3}^{1,2}\left[\bigg(\frac{\gamma}{\lambda\bar{\gamma}_{U,R}}\bigg)\left|\begin{matrix}(0,1)(1,1)\\
	(1,1)(0,1)(1,1)\end{matrix}\right.\right]\Bigg)-\frac{(1-w)}{\Gamma{[a]}\Gamma{[\mu]}}\Bigg(H_{1,2}^{1,1}\left[\bigg(\frac{\gamma}{b\bar{\gamma}_{U,R}}\bigg)^{c}\left|\begin{matrix}(1,1)\\
	(a,1)(0,1)\end{matrix}\right.\right]\\
\times \frac{1}{\gamma}H_{2,3}^{1,2}\left[\bigg(\frac{\mu\gamma}{\bar{\gamma}_{R,D}}\bigg)\left|\begin{matrix}(0,1)(1,1)\\
	(\mu,1)(0,1)(1,1)\end{matrix}\right.\right]+H_{1,2}^{1,1}\left[\bigg(\frac{\mu\gamma}{\bar{\gamma}_{R,D}}\bigg)\left|\begin{matrix}(1,1)\\
	(\mu,1)(0,1)\end{matrix}\right.\right]\\
\times \frac{1}{\gamma}H_{2,3}^{1,2}\left[\bigg(\frac{\gamma}{b\bar{\gamma}_{U,R}}\bigg)^{c}\left|\begin{matrix}(0,1)(1,1)\\
	(a,1)(0,1)(1,c)\end{matrix}\right.\right]\Bigg).
	\end{dmath}
	\hrulefill
	\vspace*{4pt}
\end{figure*}

The function $\ln{(1+\gamma)}$ can be written in terms of Fox's H-function by utilizing [Eq. (8.4.6)] in Ref.~\cite{15} and then [Eq. (1.1.2)] in Ref.~\cite{16} as
\begin{equation}
\begin{aligned}
\ln{(1+\gamma)}=H_{2,2}^{1,2}\left[\gamma\left|\begin{matrix}(1,1)(1,1)\\
(1,1)(0,1)\end{matrix}\right.\right].
\end{aligned}
\end{equation}
Now, substituting Eqs. (22) and (23) into Eq. (21) and using [Eq. (2.25.1)] in Ref.~\cite{15} and then [Eq. (2.3)] in Ref.~\cite{17} while taking into account that $\alpha=2$, we get the ergodic capacity in closed-form expression by Eq. (\ref{capacity}) in the top of next page, where $H_{.,.:.,.:.,.}^{.,.:.,.:.,.}\left[\begin{smallmatrix}.\cr .\end{smallmatrix}\Bigg\vert\begin{smallmatrix}\ .,.\cr .,.\end{smallmatrix}\Bigg\vert\begin{smallmatrix}\ .,.\cr .,.\end{smallmatrix}\Bigg\vert \psi,\chi \right]$ is the Extended Generalized Bivariate Fox's H-Function (EGBFHF) defined in~\cite{18}.

\begin{figure*}[!t]
	\normalsize
	\begin{dmath}\label{capacity}
		C=\frac{1}{\ln{(2)}}\Bigg(H_{4,5}^{3,3}\left[\frac{\mu}{\bar{\gamma}_{R,D}}\left|\begin{matrix}(0,1)(1,1)(0,1)(1,1)\\
			(\mu,1)(0,1)(1,1)(0,1)(0,1)\end{matrix}\right.\right]+wH_{4,5}^{3,3}\left[\frac{1}{\lambda\bar{\gamma}_{U,R}}\left|\begin{matrix}(0,1)(1,1)(0,1)(1,1)\\
			(1,1)(0,1)(1,1)(0,1)(0,1)\end{matrix}\right.\right]\\
		+\frac{(1-w)}{\Gamma{[a]}}H_{4,5}^{3,3}\left[\bigg(\frac{1}{b\bar{\gamma}_{U,R}}\bigg)^{c}\left|\begin{matrix}(0,1)(1,1)(0,1)(1,1)\\
			(a,1)(0,1)(1,c)(0,1)(0,1)\end{matrix}\right.\right]-\frac{w}{\Gamma{[\mu]}}\Bigg\{\\
		\times H_{2,2:1,2:2,3}^{0,1:1,1:1,2}\left[\begin{matrix}(0,0)(0,1)\cr (0,0)(1,1)\end{matrix}\Bigg\vert\begin{matrix}(1,1)\cr (1,1)(0,1)\end{matrix}\Bigg\vert\begin{matrix}(0,1)(1,1)\cr (\mu,1)(0,1)(1,1)\end{matrix}\Bigg\vert{\frac{1}{\lambda\bar{\gamma}_{U,R}},{{\frac{\mu}{\bar{\gamma}_{R,D}}}}}\right]\\
		+ H_{2,2:1,2:2,3}^{0,1:1,1:1,2}\left[\begin{matrix}(0,0)(0,1)\cr (0,0)(1,1)\end{matrix}\Bigg\vert\begin{matrix}(1,1)\cr (\mu,1)(0,1)\end{matrix}\Bigg\vert\begin{matrix}(0,1)(1,1)\cr (1,1)(0,1)(1,1)\end{matrix}\Bigg\vert{\frac{\mu}{\bar{\gamma}_{R,D}},{{\frac{1}{\bar{\lambda\gamma}_{U,R}}}}}\right]\Bigg\}-\frac{(1-w)}{\Gamma{[a]}\Gamma{[\mu]}}\Bigg\{\\
		\times H_{2,2:1,2:2,3}^{0,1:1,1:1,2}\left[\begin{matrix}(0,c)(0,1)\cr (0,c)(1,1)\end{matrix}\Bigg\vert\begin{matrix}(1,1)\cr (a,1)(0,1)\end{matrix}\Bigg\vert\begin{matrix}(0,1)(1,1)\cr (\mu,1)(0,1)(1,1)\end{matrix}\Bigg\vert{\bigg(\frac{1}{b\bar{\gamma}_{U,R}}\bigg)^{c},{{\frac{\mu}{\bar{\gamma}_{R,D}}}}}\right]\\
		+ H_{2,2:1,2:2,3}^{0,1:1,1:1,2}\left[\begin{matrix}(0,0)(0,1)\cr (0,0)(1,1)\end{matrix}\Bigg\vert\begin{matrix}(1,1)\cr (\mu,1)(0,1)\end{matrix}\Bigg\vert\begin{matrix}(0,1)(1,1)\cr (a,1)(0,1)(1,c)\end{matrix}\Bigg\vert{\frac{\mu}{\bar{\gamma}_{R,D}},{{\bigg(\frac{1}{b\bar{\gamma}_{U,R}}}\bigg)^{c}}}\right]\Bigg\}\Bigg)~\text{bps/Hz}.
	\end{dmath}
	\hrulefill
	\vspace*{4pt}
\end{figure*}

\section{Simulation and Numerical Results}
In this section, the outage, asymptotic probabilities, ASER, and ergodic capacity are verified by Monte-Carlo simulations. Furthermore, the impacts of different system parameters are investigated, for instance, Bubbles Level (BL), Water turbulence, Water type, and $\alpha-\mu$ values. BPSK is used as the modulation scheme for the ASEP simulations. \par

\begin{figure}[!t]
	\centering		\includegraphics[height=6.3cm, width=7.85cm]{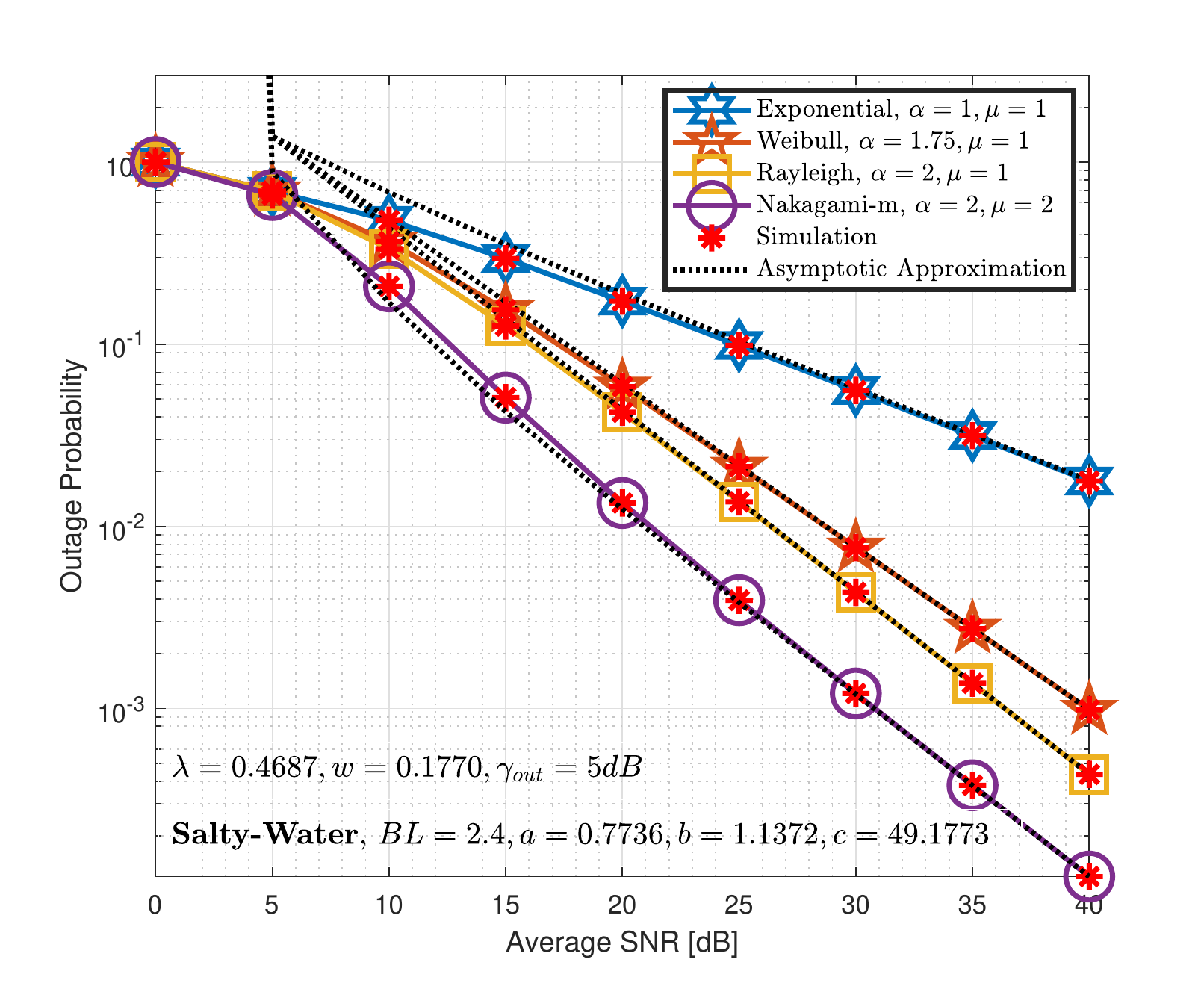}		\caption{Outage probability versus $\bar{\gamma}$ for weak optical turbulence and different values of $\alpha-\mu$.}\label{Fig.2}
\end{figure}

Fig. 2 shows the outage performance over different RF channels, e.g., Rayleigh fading. We can notice the exact matching of the analytical and asymptotic (at high SNR) expressions with Monte-Carlo simulation. This figure is produced under salty water condition with weak under-water optical turbulence (UOT). Additionally, we can notice that the RF link is dominating the outage performance by changing the parameters values of $\alpha$ and $\mu$ because the $G_{d}^{T_{3}}\ll G_{d}^{T_{2}}$ as seen in Table II. \par

The influence of under-water optical turbulence (UOT) is investigated in Fig. 2 over Nakagami-m, Rayleigh, and Weibull channel models. Again, we can observe the high degradation in the outage of almost 10 dB coding loss in Nakagami-m and 5 dB coding loss in Rayleigh channel. In this case, both UWO and RF are dominating the outage because of the equality in their diversity orders ($G_{d}$) seen as: $G_{d}^{T_{3}/T_{1}}\simeq G_{d}^{T_{2}}$ and hence, highest coding loss ($G_{c}$) is achieved as shown in Table II.\par
 
\begin{figure}[!t]
	\centering		\includegraphics[height=6.3cm, width=7.85cm]{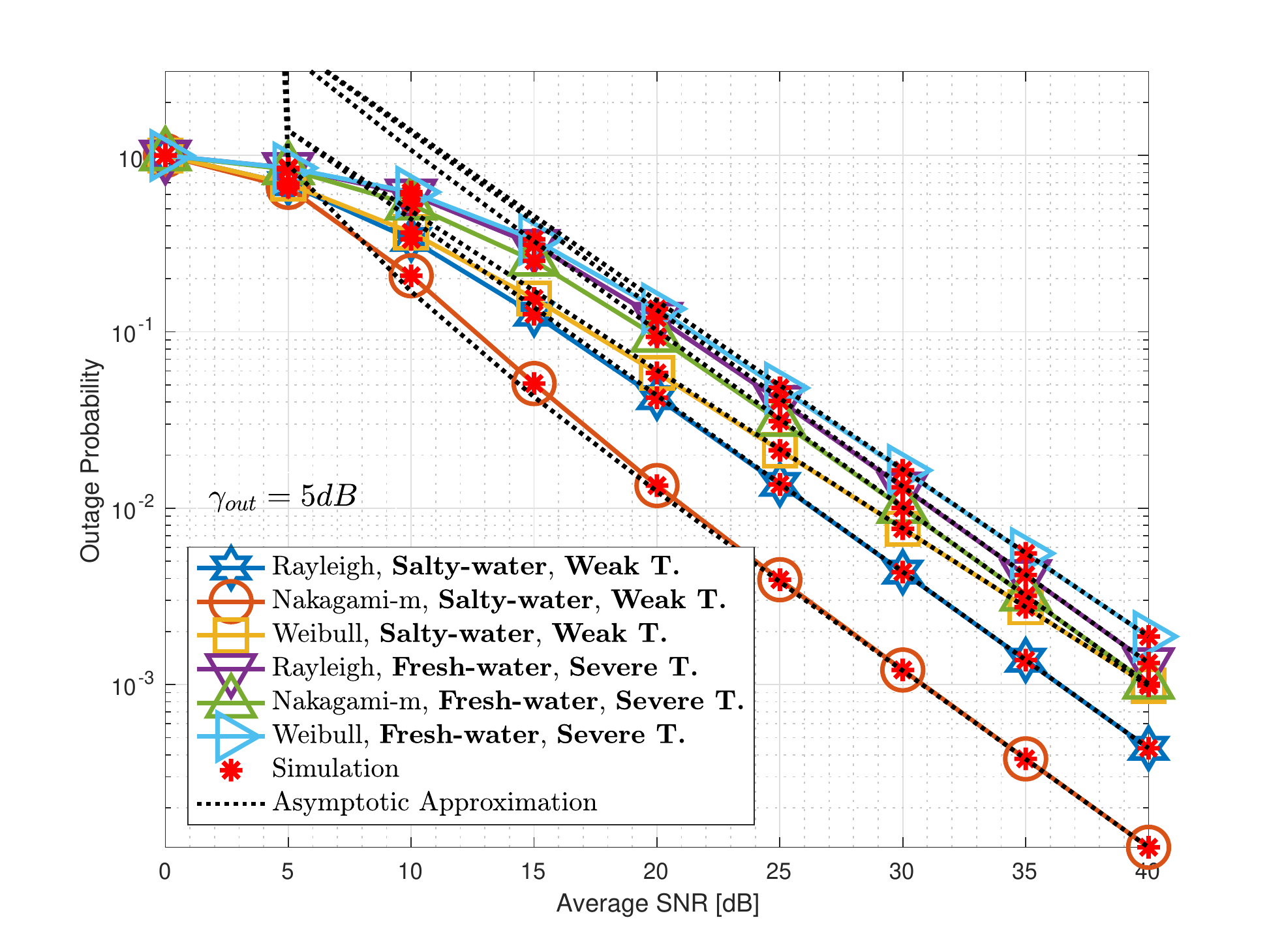}		\caption{Outage probability versus $\bar{\gamma}$ for different under-water optical turbulence conditions.}\label{Fig.3}
\end{figure}

Fig. 4 illustrates the ASER for different RF channel models. The figure is produced under salty-water with weak underwater optical turbulence condition. It can be noticed from this Fig. 4 that the RF link is dominating the performance by changing the values of $\alpha$ and $\mu$ and hence, the diversity order $G_{d}^{RF}$ is affected. Furthermore, we can see that the Nakagami-m channel has the best ASE performance while the Exponential channel is the worst  compared to others. In Fig. 5, the impact of under-water optical turbulence is investigated where we can observe the high error rate caused by UOT, e.g., Bubble levels.

Fig. 6 shows the ergodic capacity for different under-water optical turbulence conditions under heterodyne detection technique ($r=1$). We can see that when the air bubbles decreases, the ASE improves and highly achievable rate achieved and hence, better overall performance.  

\begin{figure}[!b]
	\centering		\includegraphics[height=6.3cm, width=7.85cm]{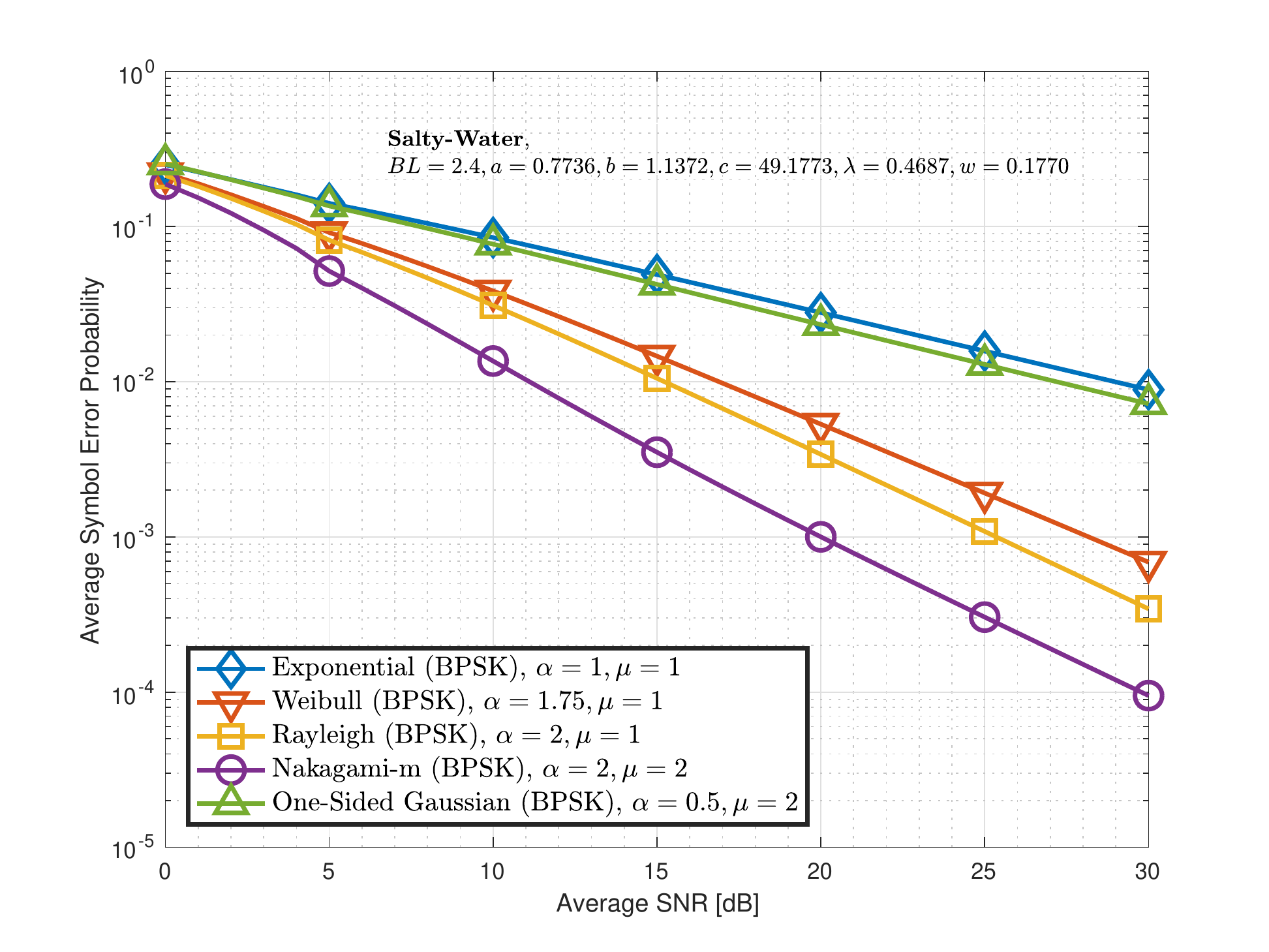}		\caption{ASER versus $\bar{\gamma}$ for weak optical turbulence and different RF channel models.}\label{Fig.4}
\end{figure}

\begin{figure}[!t]
	\centering		\includegraphics[height=6.3cm, width=7.85cm]{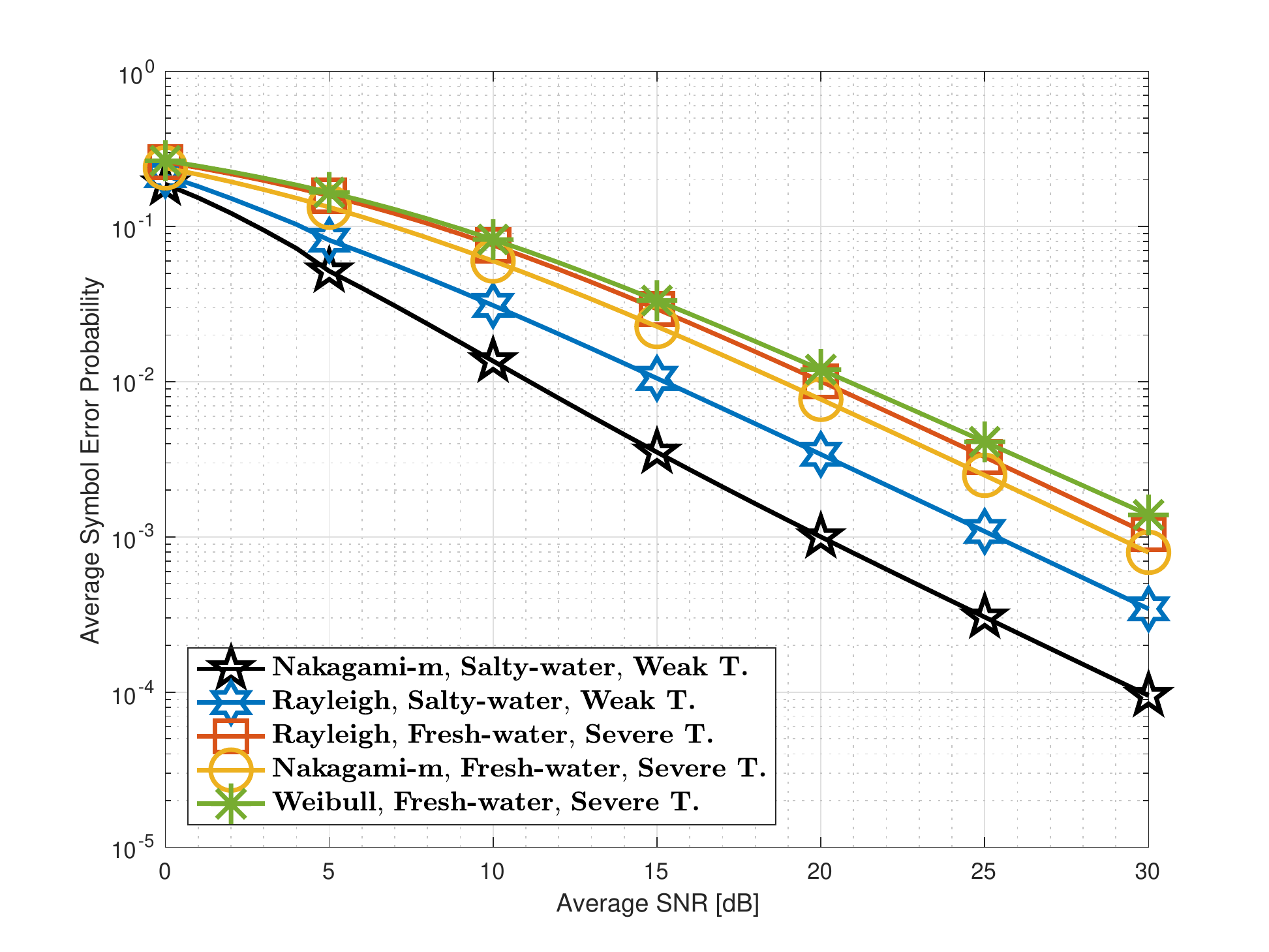}		\caption{ASER versus $\bar{\gamma}$ for different under-water optical turbulence conditions.}\label{Fig.5}
\end{figure}

\begin{figure}[!b]
	\centering		\includegraphics[height=6.3cm, width=7.85cm]{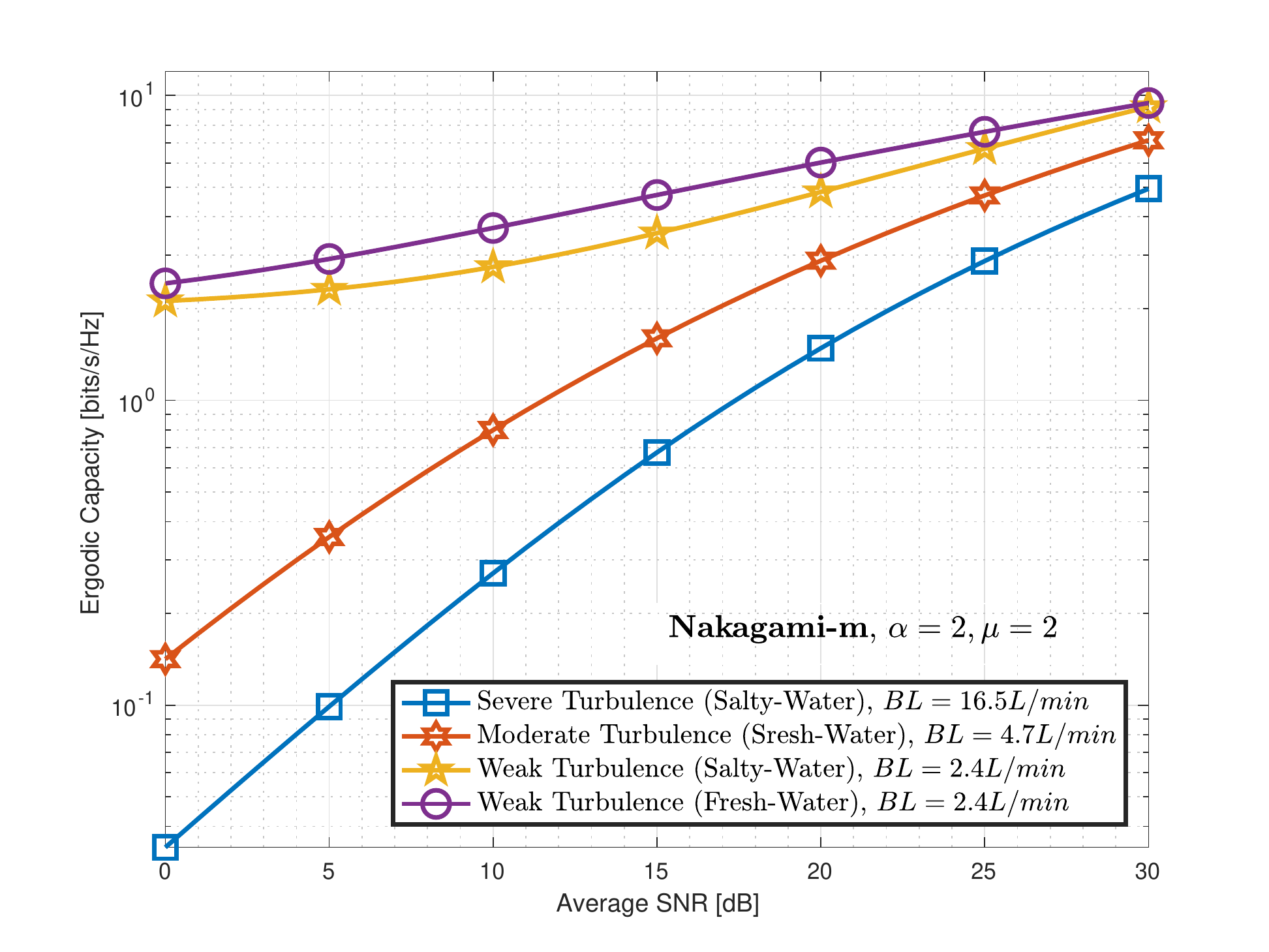}		\caption{Ergodic Capacity versus $\bar{\gamma}$ for different under-water optical turbulence conditions.}\label{Fig.6}
\end{figure}

\section{Conclusion}
In this paper, the performance analysis of a OWR mixed UWO/RF system was studied where closed-form expressions of the end-to-end outage probability, average symbol error rate, and ergodic capacity are derived. In addition, the asymptotic outage analysis is obtained for more performance insights. Exponential-Generalized Gamma (EGG) fading distribution is adopted for the first time in designing of UWOC systems; which include the effect of various impairments such as air bubbles and water salinity. Moreover, the mathematical tractability for analyzing wide range of UWOC systems. Furthermore, results show that relaying systems have a good potential for many applications, e.g., Navigation, due to the coverage area expansion and high-speed underwater communications.


%

\section*{Acknowledgment}

The  authors  acknowledge  King  Fahd  University  of  Petroleum  and  Minerals  (KFUPM)  for supporting this research.

\ifCLASSOPTIONcaptionsoff
  \newpage
\fi



%

%

\begin{IEEEbiography}[{\includegraphics[width=1in,height=1.375in,clip,keepaspectratio]{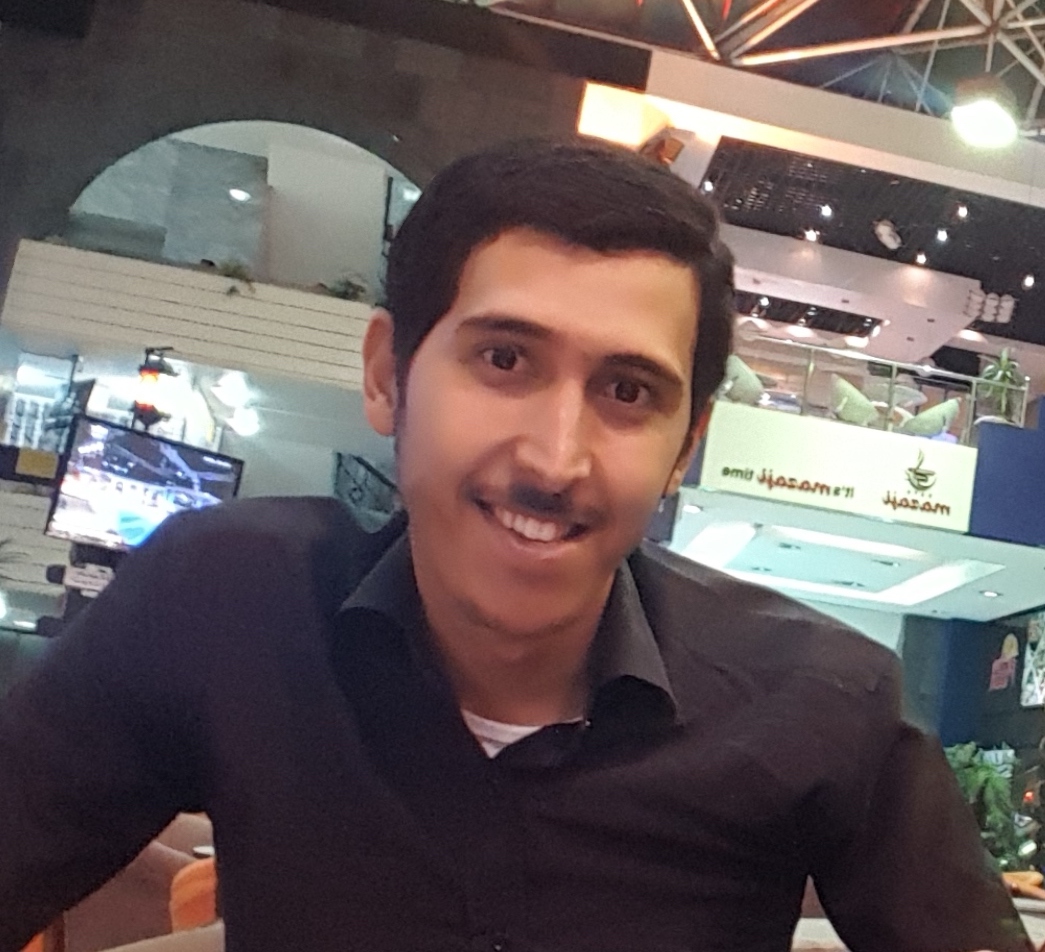}}]{Mohammed Amer}
was born in Sana'a, Yemen. He received a B.Sc. degree in Electrical Engineering from Hail University, Hail, Saudi Arabia, in 2015 (first honor) and working toward his M.Sc. degree in telecommunications engineering from King Fahd University of Petroleum and Minerals (KFUPM), Dhahran, Saudi Arabia. His research interests are design and performance analysis of wireless communication systems. 
\end{IEEEbiography}

\begin{IEEEbiography}[{\includegraphics[width=1in,height=1.25in,clip,keepaspectratio]{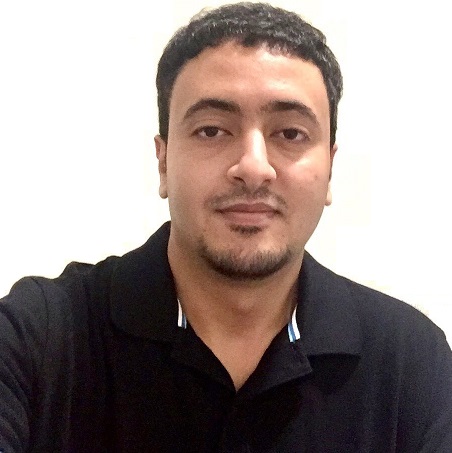}}]%
	{Yasser Al-Eryani}
	was born in Sana’a, Yemen. He received a B.Sc. degree in Electrical Engineering from IBB University, Ibb, Yemen, in 2012 and received a M.Sc. degree in telecommunications engineering from King Fahd University of Petroleum and Minerals (KFUPM), Dhahran, Saudi Arabia, in 2015. 
	He is now working towards his Ph.D. degree in electrical engineering at the University of Manitoba, Winnipeg, Canada. 
	His research interests are design, optimization and analysis of wireless communication networks.
\end{IEEEbiography}





\end{document}